\documentclass{emulateapj}
\begin{document}

\title{Photoionization Modeling and the K Lines of Iron}

\author{T.R. Kallman\altaffilmark{1} and P. Palmeri\altaffilmark{1,2}}

\affil{Laboratory for High Energy Astrophysics, NASA/GSFC}

\author{M.A. Bautista\altaffilmark{3} and C. Mendoza\altaffilmark{3}}

\affil{Centro de F\'{\i}sica, IVIC  Caracas, Venezuela}

\and

\author{J. H. Krolik\altaffilmark{4}}

\affil{Department of Physics and Astronomy, Johns Hopkins University}

\altaffiltext{1}{NASA Goddard Space Flight Center, Code 662, Greenbelt, MD 20771, USA}

\altaffiltext{2}{Department of Astronomy, University of Maryland, College Park MD 20772}
\altaffiltext{3}{Centro de F\'{\i}sica, Instituto Venezolano de Investigaciones
Cient\'{\i}ficas (IVIC), PO Box 21827, Caracas 1020A, Venezuela}

\altaffiltext{4}{Department of Physics and Astronomy, Johns Hopkins University, Homewood Campus, Baltimore MD 21218}

\date{Received ; Accepted }


\begin{abstract}
We calculate the efficiency of iron K line emission and iron K absorption in photoionized 
models using a new  set of atomic data.  These data are more comprehensive 
than those previously applied to the modeling of iron K lines from photoionized gases, 
and allow us to systematically examine the behavior of the properties of  line emission 
and  absorption as a function of the ionization parameter, density and column density of 
model constant density clouds.  We show that, for example, the net fluorescence yield for 
the highly charged ions is sensitive to the level population distribution produced by 
photoionization, and these yields are generally smaller than those predicted assuming 
the population is according to statistical weight.   We demonstrate that the effects of 
the many strongly damped resonances below the K ionization thresholds conspire to 
smear the edge, thereby potentially affecting the astrophysical interpretation of absorption
features in the 7-9 keV energy band.  We show that the centroid of the ensemble of K$\alpha$ 
lines, the K$\beta$ energy, and the ratio of the K$\alpha_1$ to K$\alpha_2$ components are 
all diagnostics of the ionization parameter of our model slabs.

\end{abstract}

 \keywords{atomic data -- atomic processes -- line
formation -- X-rays: spectroscopy}


\section{Introduction}
Iron K lines are of indisputable importance in astronomical X-ray spectra.  
They are unique among commonly observed X-ray lines in that they can be emitted efficiently by gas 
over a wide range of temperatures and ionization states.  Their location 
in a relatively unconfused spectral region gives impetus to their use as 
plasma diagnostics. They were first reported in the
rocket observations of the supernova remnant Cas A \cite{ser73},
in X-ray binaries \cite{san75,bec77} and in clusters of galaxies \cite{ser77},
the latter revealing the presence of extra galactic nuclear processed
material.  With the advent of orbiting X-ray detectors they were 
observed from almost all classes of astronomical sources 
detected in the 5-10 keV energy band.  
Further impetus for the study of these lines comes from observations of 
Seyfert galaxies and galactic black hole candidates, some of which show 
relativistically broadened and red-shifted lines
attributed to formation within a few gravitational radii of a black hole \cite{tan95}.

Recent improvements in the spectral capabilities and sensitivity
of satellite-borne X-ray telescopes ({\em Chandra}, {\em XMM--Newton})
have promoted the role of Fe~K lines as diagnostics, a trend that will continue with the
launch of future instruments such as {\em Astro-E2} and {\em Constellation-X}.
Such plasma diagnostics ultimately rely on the knowledge of the micro physics of
line formation and hence on the accuracy of the atomic data.  In spite of the
line identifications by \cite{see86} in solar flare spectra and the laboratory
measurements of \cite{bei89,bei93}, \cite{dec93} and \cite{dec95,dec97},
the K-vacancy level structures of Fe ions remain incomplete as can be concluded
from the critical compilation of \cite{shi00}. With regards to the radiative and Auger
rates, the highly ionized members of the iso-nuclear sequence, namely
Fe~{\sc xviii}--Fe~{\sc xxv}, have received much attention \cite{jac89}, and the
comparisons by \cite{chen86} and \cite{kat97} have brought about
some degree of data assurance. For Fe ions with electron occupancies greater than 9,
\cite{jac80} and \cite{jac86} have carried out central field calculations
on the structure and widths of various inner shell transitions, but these
have not been subject to independent checks and do not meet current
requirements of level-to-level data and the needs for spectroscopic accuracy 
implied by recent and future astronomical instruments.

The spectral modeling of K lines also requires accurate knowledge of
inner shell electron impact excitation rates and, in the case of photoionized
plasmas, of partial photoionization cross sections leaving the ion in
photoexcited K-vacancy states. In this respect, \cite{pal02} have
shown that the K-threshold resonance behavior is dominated by
radiative and Auger damping which induce a smeared edge.  That is, 
photons at energies below the threshold for K shell ionization can excite 
from the ground state into np (3p, 4p, etc.) K-vacancy states.  
These then decay with a high probability by spectator Auger channels, 
in which the K vacancy is filled, and an ionization occurs, by 
electrons from the  L shell.  The excitation is a resonance 
with a width determined by the Auger and radiative lifetimes
of the upper level, which can be quite short.  This leads to multiple broad and 
overlapping resonances below the K edge, with combined strength which 
equals the K photoionization cross section.   Spectator Auger
decay has been omitted from many previous close-coupling calculations of high-energy
continuum processes in Fe ions \cite{ber97,donnelly00,ber01,bal01}.  
An exception is the recent $R$-matrix computation of electron excitation rates 
of Li-like systems by \cite{whi02} where it is demonstrated that Auger damping 
is important for low-temperature effective collision strengths.

The present report is part of  a project to systematically
compute atomic data sets for the modeling of the Fe K
spectra. The emphasis of this project is on both accuracy and 
completeness.  For this purpose we make use of several state-of-the-art atomic
physics codes, together with experimentally measured line wavelengths when available,
to deliver for the K shell of the entire Fe isonuclear sequence: 
energy levels; wavelengths, radiative and Auger rates, electron impact
excitation and photoabsorption cross sections. Most of this work has already 
been reported in a series of papers,  organized loosely in order of 
descending ionization state:  energy levels, transition probabilities and 
photoionization cross sections for Fe~{\sc xxiv} were reported by \cite{bau03} 
(hereafter paper~1); energy levels and transition probabilities for the 
rest of the 'first row ions' Fe~{\sc xviii}--Fe~{\sc xxiii} were reported by 
\cite{pal02b} (hereafter paper~2); energy levels and transition 
probabilities for the 'second row ions' Fe~{\sc x}--Fe~{\sc xvii} were 
reported by \cite{men04} (hereafter paper~3); properties of the 
photoionization cross sections and the role of damping by spectator 
Auger resonances were presented by \cite{pal02} (hereafter paper~4); 
energy levels and transition probabilities for the 'third row ions' Fe~{\sc ii}--Fe~{\sc ix}
were reported by \cite{pal03} (hereafter paper~5); photoionization and electron impact 
cross sections were presented for the first row by \cite{bau03} (hereafter paper~6); 
and photoionization cross sections for the second row by \cite{bau04} 
(hereafter paper~7). 

In the present paper we present a synthesis of all the data  
in the series and illustrate the application of these data to calculations of 
opacity and emission spectra of equilibrium photoionized plasmas.  We explore the systematic 
behavior of the emission line profile shapes as a function of ionization 
parameter, the line emissivity as a function of gas density, column density and optical depth,
and the absorption line and continuum properties as a function of ionization parameter.
We show that the line profiles, although rich with detail, obey certain systematic 
behaviors.   The K shell opacity also contains many features which depend on the ionization 
balance in the gas.  Much  of the data used for iron K shell which had been 
incorporated into previous models for the X-ray spectrum is shown to be overly simplified, and 
in some cases incomplete.  All of the ingredients and results of the calculations published 
here are publicly available as part of the {\sc xstar} photoionization code.

The models presented here are simple uniform slab models, but include an extensive 
suite of atomic data together 
with self-consistent ionization, excitation, and thermal balance.  The resulting
synthetic spectra are somewhat idealized, since they omit any density or pressure gradients
and assume normal incidence for the illuminating radiation.  They also use an approximate
treatment of continuum Compton scattering, which is likely to be 
important at the large columns where iron K is emitted efficiently.  Therefore 
they serve primarily as illustrations 
of the relative importance of various physical effects.  These include the 
relative prominence of spectral features associated with damped lines, 
both in emission and absorption, and dependence of these quantities
on ionization parameter and optical depth.   More realistic synthetic spectra, 
involving accurate Comptonization in static atmospheres and winds, 
will be presented in a subsequent paper.

\section{Model Ingredients}

\subsection{Atomic Data}

The atomic data used in this are compiled from the data sets reported in the 
previous papers in the series, so we provide here only a brief summary of the 
assumptions and procedures used in the calculations.  For the purposes of 
discussion, both here and elsewhere, the ions in the iron isonuclear sequence
can be loosely classified according to the row in the periodic table associated 
with the various isoelectronic sequence.  We emphasize, here and in what follows, that
our goal is the treatment of the transitions and spectral features associated with the 
K shell of iron, so we have greatly simplified and omitted much of the physics associated 
with the outer shells in our calculations.  We are confident that this does not 
significantly affect the accuracy of the results for the K shell, and we discuss 
reasons for this in what follows.

Computation of the atomic parameters for the iron ions requires taking into account
the electron correlation and the relativistic corrections together. In our different
studies detailed in papers~1--7, the approach we took is based on the use and comparison
of results of several computational platforms, all of them publicly available, 
namely {\sc autostructure} \cite{bad97}), 
{\sc hfr} \cite{cow81}, and the Breit--Pauli R-matrix package ({\sc bprm}) based on the close-coupling
approximation of \cite{bursea71}. While the latter was preferred for calculating
the electron and photon scattering properties due to its accurate inclusion of the continuum
channel coupling, the former two were chosen for their semi-empirical options which
can utilize experimentally measured  energy level spacings to improve 
the theoretical level separations for which the inter-system transition
rates are sensitive. Although all these codes include relativistic effects and configuration
interactions, {\sc autostructure} has the most complete Breit Hamiltonian, i.e. the two-body part
of the magnetic interactions. The Breit interaction plays a key role in the highly charged ions of
the first row in particular for transitions with small rates and for those involving states
subject to strong relativistic couplings. 
It was found that the energies and rates are also affected by core relaxation effects
(CRE) and that these effects are more important than the higher order relativistic interactions
for the spectator inner shell processes that dominate in the second and third row ions.
Since, among the platforms used at the time this project was initiated, 
{\sc hfr} handled CRE more efficiently in ab initio
calculations ({\sc autostructure} has since been updated to treat this process efficiently), 
it was used for data production in cases where no experimental level energies
were available for the semi-empirical corrections, i.e. in the second and third row ions.

For the first row, a complete set of level energies, wavelengths, $A$-values,
and total and partial Auger rates have been computed for the K-vacancy states
within the $n=2$ complex. The experimental level energies listed in the
compilation by \cite{shi00} are used in semiempirical adjustment
procedures. It has been found that these adjustments can lead to large
differences in both radiative and Auger rates of strongly mixed levels.
Several experimental level energies and wavelengths have been questioned as a
result, and significant discrepancies are encountered with
previous decay rates. These have been attributed to numerical problems in
the older work. The statistical accuracy of the level energies and wavelengths
is ranked at $\pm 3$ eV and $\pm 2$ m\AA, respectively, and that for $A$-values
and partial Auger rates greater than $10^{13}$ s$^{-1}$ at better than 20\%.
Photoabsorption cross sections across the K edge and electron impact K-shell
excitation effective collision strengths have also been calculated. The target
models are represented with all the fine-structure levels within the $n=2$
complex, and the effects of radiation and spectator Auger dampings are
taken into account by means of an optical potential \cite{gorczyca00}.
In photoabsorption, these effects cause the resonances converging to the K
thresholds to display symmetric profiles of constant width that smear the edge
with important implications in spectral analysis. In collisional excitation,
they attenuate resonances making their contributions to the effective
collision strength practically negligible.

Regarding the second-row ions, the K$\alpha$ and KLL Auger widths
are found to be nearly independent of both outer electrons and
electron occupancy keeping a constant ratio of $1.53 \pm 0.06$.
By comparing with previous theoretical and measured wavelengths,
we estimate the accuracy of the reported wavelengths to be within
2 m\AA. Also, the good agreement found between the radiative and
Auger data sets that are computed with different platforms allow
us to propose with confidence an accuracy rating of 20\% for the
line fluorescence yields greater than 0.01. The predicted
emission and absorption spectral features seem to be in good
correlation with measurements in both laboratory and
astrophysical plasmas. For computer tractability, photoionization
cross sections across both the L and K edges 
have been computed in $LS$ coupling and later split among fine
structure levels according to the formula of \cite{rau76}. Targets are
modeled with all the singly excited terms within the $n=3$
complex. In similar fashion as for the first-row ions, the
effects of radiative and Auger dampings are taken into account.
\cite{dec95} have published the only experimental study of the
K-line spectrum of the Fe second row, and they found that the
only ion for which the K lines are sufficiently unblended to
derive accurate wavelengths is for Fe~{\sc x}.  Their wavelengths
for the doublet in the latter are approximately consistent with
the ab-initio values computed by us and by \cite{dec95}.

There are no experimental data for Fe ions of the third row other than for
Fe~{\sc ii}. For this ion, measurements have been made in the solid, and good
agreement is found with our calculations. Level energies, wavelengths,
$A$-values, Auger rates, and fluorescence yields have been obtained
for the lowest fine-structure levels populated by photoionization of the
ground state of the parent ion. Due to the complexity and size of
level-to-level Auger calculations, we have employed a compact formula
to compute Auger widths from {\sc hfr} radial integrals \cite{pal01}.
An accuracy of $\sim$10\% is estimated for the transition probabilities as a
result of a comparison of K$\beta$/K$\alpha$ branching ratios
taken from different theoretical and experimental data sets in the literature.
Concerning the photoabsorption cross sections, a model similar to
the one used in Paper~4 has been constructed whereby a single
1s~$\rightarrow$~$n$p Rydberg series of Lorentzian resonances converging to
each K threshold has been considered in each ionic species. The partial
photoionization cross sections connecting the ground state of the parent ions
to the different K-vacancy states have been calculated by {\sc autostructure}
in a single-configuration approximation.

In the present self-consistent photoionization model, we have incorporated
all our published K-vacancy levels for the first-row ions and those populated
by photoionization of the parent-ion ground state according to the selection
rules of \cite{rau76} in the case of species of the second and third rows.
These are the levels whose populations are calculated explicitly and for
which the transition rates given in our previous papers have been adopted. In
addition superlevels are included that are associated with the photoabsorption
features near the inner-shell edges using the {\sc bprm} cross sections
of Papers~6--7. The populations of these superlevels are also calculated,
and their effects on the ionization and thermal balances are taken into account.
Superlevel decays are treated as `fake' transitions, i.e. lines with
wavelengths much longer than any of interest to the model results.
Such fake transitions are also used to treat the decays of other levels
for which we do not explicitly treat the line wavelengths such as the
decays of L-vacancy levels in the second and third row ions.
Figure~1 shows two examples of energy-level diagrams: in Fe~{\sc xviii},
a typical case of a first row ion, and in Fe~{\sc x} which is
representative of a second-row ion.
In Fe~{\sc xviii}, the $[{\rm 1s}]{\rm 2p}^6\ ^{2}{\rm S}_{1/2}$ K-vacancy
state is populated by photoionization of the ${\rm 2p}^6\ ^{1}{\rm S}_0$
ground state of Fe~{\sc xvii}. It then decays to or can be populated from
the ground term ${\rm 2p}^5\ ^{2}{\rm P}^{\rm o}$ through K$\alpha$ transitions.
It can also decay via KLL Auger transitions or be populated by the inverse
process, i.e. KLL dielectronic recombination from the ground configuration
${\rm 2p}^4$ of Fe~{\sc xix}. Photoabsorption near the K threshold is
considered through a transition from the ground state to the [K] superlevel.
That is, the full complexity of the energy dependent absorption cross section
is calculated at all energies, but the portion below the K ionization threshold
is used to calculate the photoexcitation rate into the [K] superlevel rather
than as direct ionization. This $[K]$ superlevel can then decay to the ground or
autoionize.

In the case of Fe~{\sc x}, the $[{\rm 1s}]{\rm 3p}^6\ ^{2}{\rm S}_{1/2}$
K-vacancy state is populated by photoionization of the
${\rm 3p}^6\ ^{1}{\rm S}_0$ ground state of Fe~{\sc ix}. It then decays either
to the ${\rm 3p}^5\ ^{2}{\rm P}^{\rm o}$ ground term through the
K$\beta$ channel or to the
$[{\rm 2p}]{\rm 3p}^6\ ^{2}{\rm P}^{\rm o}_{1/2,3/2}$ L-vacancy states
via the K$\alpha$ channel. It can also decay through Auger transitions, but
here we neglect the dielectronic recombination channels from the
${\rm 3p}^4$ ground configuration of Fe~{\sc xi} because they involve only
KMM autoionization rates which are orders of magnitude weaker than those of
the main KLL Auger decay channels. We do not treat the cascade from L-vacancy
states, and therefore the $[{\rm 2p}]{\rm 3p}^6\ ^{2}{\rm P}^{\rm o}_{1/2,3/2}$
levels are connected to the ground levels with fake transitions.
Photoabsorption near the K and L thresholds is considered via transitions
that connect the [K] and [L] superlevels to the ground level.  That is, we
take the full absorption cross section including resonance structures from
the $R$-matrix calculations, and attribute this part of it to photoexcitation
into these superlevels. The photoionization transition to the [L]
superlevel is associated with small contributions of L-shell channels to the
cross section near the K edge. These contributions cascade through Auger
decays to the ground level of the neighboring ion, here Fe~{\sc xi}. In the
third row, the K-vacancy states decay to L-vacancy states (K$\alpha$ lines)
and to 3p subshell vacancy states (K$\beta$ lines). The former are populated
exclusively by photoionization from the ground level of the parent ion. We
have applied the same treatment of the cascade as for the second-row ions.
The L-vacancy and 3p-vacancy levels decay to the ground level through fake
transitions. The photoabsorption near the K edge is described in terms of a
single 1s~$\rightarrow$~$n$p Rydberg series converging to the different K
thresholds.

Figures~2--3 plot the energies of the lines and K edges in our
data set as a function of ionization stage which can be compared
with those presented by \cite{house69} and by \cite{mak86}. For
each ion the energies of all the components of a line or a K
photoionization edge are plotted with equal weight and without
accounting for broadening due to damping or blending into an
unresolved transition array (UTA). So the dispersion in the line
or photoionization threshold-energy for each ion represents the
range in centroid energies of various lines or continua. This
does not take into account the broadening due to damping or the
relative strengths of the various components arising from the
distribution of upper-level populations and transition
probabilities. In Section~2.2 synthetic profiles are presented
which take these factors into account.  The distinction between
the low and high ionization stages is apparent: for the third row
the dispersion in line and K-threshold energies within a given
ion can be comparable to the difference in these quantities
between adjacent stages although the K$\beta$ line provides a
sensitive diagnostic of the charge state.  For example, the
K$\alpha$ energies are in the range  6385--6403 eV in Fe~{\sc
ii--vii}.  For Fe~{\sc viii--xi}, the K$\alpha$ energy interval
actually decreases slightly to 6381--6403 eV.  This is somewhat
counter-intuitive since generally binding energies increase with
increasing ionization causing all line energies to increase.
However, the K$\alpha$ lines are sensitive to both the 1s and 2p
electron binding energies. The former does increase with
increasing ion stage while the 2p energy decreases slightly due
to the changing interaction with the 3d shell electrons. Further
evidence for this comes from the fact that the K$\beta$ line
energy increases between Fe~{\sc i} and Fe~{\sc xi} as the
K$\alpha$ energy decreases.  For Fe~{\sc xii--xiii} the line
intervals increase to 6380--6415 eV, for Fe~{\sc xiv--xv}
increasing by approximately 3 eV per ion from 6385--6422 eV to
6391--6428 eV.  Fe~{\sc xvi} has a particularly simple structure
owing to its Na-like configuration, with lines at 6414 eV and
6426 eV. For the first row, K$\beta$ is unimportant, and there is
much greater separation between various ion stages. Each ion
stage has line energies ranging over $\sim$200 eV. Note that some
first-row ions can emit lines at energies below 6.4 keV; for
example, transitions such as ${\rm 1s2s}^2{\rm 2p}^4-{\rm
1s}^2{\rm 2p}^5$ in Fe~{\sc xix} produce lines near 1.99 \AA\
(6.23 keV). These transitions are strongly affected by
configuration interaction with, for instance, ${\rm 1s}^2{\rm
2s}^2{\rm 2p}^3$ and have relatively small $A$-values
($\sim$10$^{12}$ s$^{-1}$).

Each {\sc bprm} cross section in Papers~6--7 (first and second row ions) is
tabulated on an energy grid containing several thousand points. In order to
include these data into the {\sc xstar} database, compression is invoked using
the resonance-average photoionization (RAP) method and the numerical
representation of RAP cross sections proposed by \cite{bau98}. The choice
of the accuracy parameter ($\Delta \equiv \delta E/E $) is guided by the
instrumental resolution of the XRS on board {\em ASTRO-E2} at the energies
of the iron K lines, i.e.  $\Delta=0.0001$ (10 times the XRS resolving power
at 6.4 keV). Figure~4 plots the photoabsorption cross sections for the various
ions, and in Table~\ref{feat} we give the strongest near K-edge absorption
features that can be resolved by the XRS (resolving power of 1000). Each
feature can be due to a superposition of several resonances; the feature
parameters, namely wavelength, oscillator strength, and full width at half
maximum, are fitted values extracted from Gaussian profiles using RAP cross
sections with $\Delta=0.001$.

Examination of Figure~4 shows a variety of different cross-section behaviors,
depending crudely on ionization stage (i.e. first, second, or third row)
and on the fine structure of both the valence- and inner-shell electron
configurations. In the third row, resonance features are close to the
threshold often overlapping and K$\alpha$ and K$\beta$ do not appear.
The second row shows more complex behavior, and K$\beta$ also appears in
absorption. In the first row both K$\alpha$ and K$\beta$ appear in absorption,
along with a very rich and distinct resonance spectrum below threshold.
More details of the resonance features and a description of the level scheme
adopted for each ion are given in the Appendix.

\subsection{\sc xstar}

Results in the following section were calculated by incorporating the 
K vacancy energy levels, and the transitions connecting them to other 
states of the ions of iron, into the {\sc xstar} 
modeling code (Kallman and Bautista 2001). {\sc xstar} calculates the 
level populations, ion fractions, temperature, opacity and emissivity 
of a gas with specified elemental composition, under the assumption 
that all relevant physical processes are in a steady state.  
The calculation is a full collisional-radiative model in the sense that 
all level populations are explicitly calculated, and the LTE balance 
is attained under the conditions of high gas density or Planckian
radiation field.  Radiative equilibrium is achieved by calculating 
the integral over the net emitted ($C$) and absorbed energy ($H$) 
in the radiation field and varying the gas temperature until the integrals satisfy 
the criterion $(H-C)/(H+C) \leq 10^{-4}$.
Radiative transfer of the ionizing continuum is treated with a simple two stream
approximation, and diffusely emitted lines and continua 
are subject to trapping according to an escape probability formalism.  
Expressions for escape probabilities are given in \cite{kal01}, 
and these are applied uniformly to all emitted photons using optical 
depths which are calculated self-consistently at each point in the model slab
based on the database oscillator strengths, Voigt profiles 
accounting for thermal Doppler broadening and 
natural broadening, and the column density of the relevant lower level 
integrated over the slab.  
{\sc xstar} by itself does not allow accurate treatment of elastic scattering 
due to Compton scattering.  This process is treated in the same way as absorption, 
which will tend to overestimate the attenuation at large column densities, 
$N \geq 10^{24}$ cm$^{-2}$.  Models which incorporate a more 
accurate treatment will be presented in a subsequent paper.  

In the models presented here we have not included the effects of 
resonant photoexcitation on the line emission.  This process affects K$\alpha$ lines only 
from the first row ions, and it depends on the geometry assumed for the 
line emitting gas:  if the emitting gas is spherically symmetric surrounding the 
continuum source, and stationary, then photoexcitation is negligible.  
Fluorescence emission is not affected by this assumption. 

Previous versions of {\sc xstar} had a limited treatment of the iron K shell, 
utilizing the photoionization cross sections of \cite{ver86} for 
the K shell photoionization, together with the line energies and fluorescence
and Auger yields from \cite{kaa93}.  In order to incorporate the 
new atomic data rates we have added our chosen subset of the K vacancy levels 
and line transitions connecting them with L- and M-shell levels.  We account for 
the full radiative-Auger cascade following a K-shell photoionization event in any
first row ion. For the intermediate and low charged ions, i.e. the second
and third row ions, multiple ionization following the ejection of a K-shell photo electron
is incorporated by adding non-radiative decays of the K-vacancy levels into the ground
levels of the neighboring  ions.  
The rates used for these transitions are obtained by multiplying the Auger widths of the 
K-vacancy upper level by the branching ratios given by \cite{kaa93}.
In these cases, we do not treat L- and M-fluorescence line emission 
during Auger cascade, since our atomic calculations do not include decays of 
L shell vacancy levels.  This omission has a small effect on our 
ionization balance since the rates for valence shell photoionization always 
exceed the rate for K shell ionization by large factors.

The treatment of K shell photoionization and the associated opacity is divided into 
two parts.  Absorption events which result in direct ionization, i.e. photons 
with energies above the threshold for photoionization, are treated using 
the conventional expression (eg., Osterbrock 1978)  integrating
the product of the photoionization cross section and ionizing photon flux over energy.  
Absorption by photons just below threshold are treated differently depending on  the type
of ion, i.e. first, second, or third row ion. For a first row ion, the resonant photoabsorption
events that connect the ground state to excited $[{\rm 1s}]n{\rm p}$ states lying just below
the K threshold are lumped together into a superlevel [K] which decays back to the
ground state through a fake radiative transition.  The rate for excitation into these levels, and 
the associated opacity, are treated in a manner analogous to photoionization, i.e. with a
continuous opacity distribution convolved with the radiation field.  
In the second row, another superlevel is
added, superlevel [L], in order to incorporate the photoabsorption structure near the L edge.
The contribution of the L-shell photoionization to the K-shell photoabsorption cross section
near the K threshold is treated as a transition from the ground level of the parent ion to
the superlevel [L] which can auto ionize to the ground level of the next charge stage.
Concerning the third row ions, the photoabsorption features near threshold are implemented 
as photoionization transitions that connect the ground level of the parent ion to the K-vacancy levels.
The photoabsorption cross section for each ion then can be divided into 
contributions from bound-bound transitions into explicitly treated levels (the K$\alpha$
and K$\beta$ lines are examples), bound-bound transitions into resonances near thresholds 
which are lumped together into a superlevel or into photoionization 
transitions into inner shell thresholds, and direct photoionization  
by the various sub shells.  

The ionization balance of iron, and temperature structure of our models, are shown
in figures~5 and 6.  These were calculated using a simple 
$F_\varepsilon \propto \varepsilon^{-1}$ ionizing spectrum and cosmic abundances 
\cite{gre98}.  As a rule of thumb, this figure shows that first row ions 
predominate for log($\xi$)$\geq$ 2, second row ions are important 1 $\leq$ log($\xi$)$\leq$ 2,
and third row ions for  log($\xi$)$\leq$ 1.

\section{Results}

\subsection{Analytic Behavior}

For the purpose of illustrating the behavior of K lines in photoionized gas, we 
consider line emission and absorption in a spherical shell of gas surrounding a point source  
of continuum radiation.  We assume cosmic element abundances \cite{gre98}, constant gas 
density, and an $F_\varepsilon \propto \varepsilon^{-1}$ power law ionizing spectrum.
In such a cloud the emissivity of the K lines is given by

$$j=\varepsilon_K \int_{\varepsilon_{Th}}^\infty{F_\varepsilon \sigma_K(\varepsilon)
{{d\varepsilon}\over{\varepsilon}}} \omega_K n y_{Fe} x_l ~~~~(1)$$

\noindent where $\varepsilon_{Th}$ is the photoionization threshold energy, $F_\varepsilon$ is the 
local flux,  $\sigma_K(\varepsilon)$ is the K shell photoionization cross section, 
$\omega_K$ is the fluorescence yield, $n$ is the gas density, $y_{Fe}$ is the iron elemental 
abundance, and  $x_l$ is the population of the lower level.  This can be rewritten in terms 
of the local specific luminosity, $f_\varepsilon={{4 \pi R^2}\over{L}} F_\varepsilon$, where $R$
is the distance from the source of ionizing radiation to the inner shell surface, and $L$ is 
the ionizing luminosity of the continuum source.  We define the emissivity per particle:

$${{j}\over{n^2}}=\varepsilon_K {{\xi}\over{4\pi}} \omega_K y_{Fe} x_l 
\int_{\varepsilon_{Th}}^\infty{f_\varepsilon \sigma_K(\varepsilon)
{{d\varepsilon}\over{\varepsilon}}} ~~~~(2)$$

\noindent and $\xi=L/nR^2$ is the ionization parameter (Tarter Tucker and Salpeter, 1969).
This shows that the line emissivity is proportional to the ionization parameter.
The line luminosity emitted by a thin spherical shell of uniform ionization, 
composition and density is then

$$L_{line}=N L  \omega_K y_{Fe} x_l 
\int_{\varepsilon_{Th}}^\infty{f_\varepsilon \sigma_K(\varepsilon)
{{d\varepsilon}\over{\varepsilon}}} ~~~~(3)$$

\noindent and the line equivalent width is

$$EW_{line}={{N \omega_K y_{Fe} x_l}\over{f_{\varepsilon K}}}
\int_{\varepsilon_{Th}}^\infty{f_\varepsilon \sigma_K(\varepsilon)
{{d\varepsilon}\over{\varepsilon}}} ~~~~(4)$$

\noindent where $N$ is the radial column density of the shell.
If $f_{\varepsilon} \propto \varepsilon^{-1}$,
$\sigma_K(\varepsilon) \propto \varepsilon^{-3}$, $x_l$=1, $\omega_K$=0.34,
and cosmic iron, 

$$EW_{line}={{N \omega_K y_{Fe} \sigma_{Th} \varepsilon_K}\over{4}}
\simeq 0.3 {\rm keV} N_{24} ~~~~(5)$$

\noindent and $N_{24}$ is the column in units of $10^{24}$ cm$^{-2}$, 
and we have adopted the hydrogenic value for the threshold cross section.

Equations (3)-(5) show that the line luminosity depends on factors related to 
atomic rates, the ionization balance, the intensity of the ionizing radiation, 
the amount of emitting gas (here parameterized by the cloud column density).
In what follows we present the dependence of the emission line properties on 
various of these factors:  ionization parameter, gas density, line optical 
depth, and column density.  

\subsection{Fluorescence Yields}

As a check on our results we have calculated fluorescence yields for each line, and for each ion averaged 
over all the K lines.  We do this by calculating the ratio of the net radiative rate out of the upper level for 
each K line by the net rate into the level.  The average over  the ion is calculated by taking the ratio of the 
total line emission rate to the total excitation rate for all the levels of the ion.  
These are summarized in table 2.  The first column is the ion stage, the second column is the average 
per-ion fluorescence yield calculated using the above procedure, which is equivalent to weighting the 
individual lines according to the excitation rate when calculating the average.  The third column 
shows the average per-ion yield calculated by assuming the levels are populated according 
to statistical weights.  This is the assumption intrinsic to the configuration average values quoted by 
most previous authors, such as \cite{jac86} and \cite{kaa93}.

Table 2 shows that the yield is generally a slowly varying function of 
the ion charge state, increasing from the classical value of 0.34 for Fe II through the third and second row, and 
then decreasing beyond Fe XIX.  Comparison of the the second and third columns shows that the 
averaging makes little difference for ions Fe II -- Fe XVII, i.e. the second and third rows of the 
periodic table.  This is due to the fact that the 
Auger and radiative probabilities are independent of the valence shell structure for the these ions, 
a point which has been made in our previous papers.  This means that the yields for all the K line upper levels
are approximately constant for these ions, although we find variations by $\simeq 20 \%$ for some 
ions.  In the ions of the first row of the periodic table, Fe XVII -- XXIV, the 
individual level yields can differ by large factors (see paper 2), and the averaging scheme 
is important.  Thus we find that for Fe XXII and Fe XXIII the average yield is considerably lower 
than that found by \cite{jac86}, owing to the fact that photoionization from the 
ground state of the parent ion tends to select K vacancy levels which have lower yields.
The value for Fe XXIV is affected by the fact that there is no dipole-allowed 
decay of the most probable final state of a K shell photoionization event. 

\subsection{Ionization Parameter Dependence}

In figure~7 we plot the ratio of the  emissivity per particle for K line production to the 
ionization parameter for a sequence of model slabs with column density $N=10^{17}$ cm$^{-2}$
as a function of ionization parameter, $\xi$.  The various curves correspond to 
the contribution from the ions of iron, summed over the K lines of each ion.

These curves were calculated using a set of {\sc xstar} models of thin spherical shells 
illuminated by the same $f_{\varepsilon} \propto \varepsilon^{-1}$ spectrum
as was used in deriving equation~(5).  Using the same fluorescence yield, cross 
section, abundance, etc., in equation~(2) as was used in equation~(5), we 
predict ${\rm log}(j/n^2) \simeq  -27.3 \xi$, while figure 7 gives 
${\rm log}(j/n^2) \simeq  -27.6 \xi^{1.2}$.  

Figure 7 shows that the emissivities of the various ions peak close to the ionization parameter 
where the corresponding ion fraction peaks.  Differences between the curves are due to 
differences in the ionization balance, K shell photoionization rate, and Auger yield.
Comparison with figure 5, the ionization distribution, shows that the contribution of a given 
ion to the line emissivity is shifted down in ionization parameter to the value where the parent 
ion dominates the ionization distribution, thus Fe XIX dominates the line emission near 
log($\xi$)=2, where Fe XVIII dominates the ionization balance.  Also, ions which appear 
prominently in the line emissivity distribution are generally not those which dominate the 
ionization balance, but rather their ionization products.  Thus Fe XVIII and XIX appear 
more prominently in the line emissivity plot than does Fe XVII, although the converse 
is true in the ionization balance.  Similar behavior occurs for Fe X vs. Fe IX.  This is 
of interest because ions such as Fe X and Fe XVIII have nearly filled atomic shells 
(K-like and F-like) and therefore simpler level structure than ions with half-filled subshells.  
This may aid in interpretation of spectra.

\subsection{Density Dependence}

Essentially all the K lines of observational interest have transition probabilities which 
are greater than 10$^{15}$ s$^{-1}$.  As a result, very high densities are 
required for collisional suppression of K lines under astrophysical conditions.  At lower 
densities collisions can  play a role in determining the populations of the 
lower levels for some K lines.  This is because many of the 
valence shells have fine structure levels with critical densities less than 10$^{12}$ cm$^{-3}$.  
The populations of these levels determine the relative strengths of the various components of the 
K lines.  These populations can also affect the dependence of line strengths on optical depth, which 
will be discussed in the following section.

Density dependence arises because of mixing of the fine structure levels of the ground term
Such mixing can either enhance or suppress the net line emission from a given ion, depending 
on whether the excited sublevels can ionize to K vacancy levels with larger or smaller 
fluorescence yields.  In either case, the spectrum will be changed, as line components of 
different energy are emitted.  Our treatment of the photoionization and excitation accounts for 
all the photoionization channels, and the fluorescence and Auger decays of the K vacancy levels 
they produce, for all the levels of the first row ions which may be of interest in 
photoionization.  That is, we consider K shell photoionization from all the levels within $\simeq$50 eV 
of the ground level, which includes all the levels of the ground terms of these ions.  For the 
second and third row, our treatment does not accurately account for 
differences in the K line excitation properties of excited levels.  That is, we 
include photoionization and the subsequent line emission processes for 
the excited levels of the ground term, but these rates are duplicates of rates for the ground level.
So for these ions we cannot accurately predict density dependent effects.  

Figure~8  shows the ratio of the line emissivity at a given density to that at a density 10$^{8}$ cm$^{-3}$ 
from a series of model slabs 
with ionization parameter log($\xi$)=2 and column density 10$^{17}$ cm$^{-2}$.  In preparing 
this figure we have also corrected for the fact that the ionization balance 
depends (weakly) on the density, by dividing the line emissivity by the fractional 
abundance of the parent ion at each density.  Since density 
acts to mix the levels of the ground term, the effect of increasing density is to redistribute the 
population from the ground level to the other levels in the term.  The net effect summed over the lines 
of a given ion is small for many ions, since the line fluorescence 
efficiency is similar for K shell ionization of the various levels of the ground configuration.
On the other hand, some ions show a strong density effect, and this can act either to enhance 
or suppress the total K line emission for that ion.  For Fe XXII, the line emission is 
enhanced at density great than $\sim 10^{12}$ cm$^{-3}$, due to mixing of 
population into the J=1 level of the ground term of 
the parent ion, Fe XXI.  This level has a larger photoionization cross section than the ground 
level, leading to enhanced line emission.  For Fe XXI,  which is produced by photoionization of Fe XX, 
the effect is opposite; the excited levels of the ground term have lower photoionization cross section 
than ground.  The magnitude of this effect is a factor $\sim$2 -- 3.  
This figure also shows onset of collisional suppression which affects all the 
K lines at density $\geq 10^{15}$ cm$^{-3}$.  

\subsection{Optical Depth Dependence}

K lines from second and third row ions are not subject to resonance scattering because 
their lower levels are inner shell vacancy states, and so do not exist in 
significant populations in astrophysical plasmas.  In first row ions 
the partially filled L shell allows resonance absorption of K line photons.  As has been pointed 
out by \cite{ros96} and others, this can lead to efficient destruction of these lines since 
there is a significant ($\geq$10$\%$) probability of Auger decay for the upper levels  each time a 
line photon scatters.  This Auger 
destruction has been suggested as an explanation for the apparent absence of K lines from first 
row ions in some astrophysical sources.  On the other hand, the existence of multiple lower levels 
for the K lines from many of these ions allows for alternate radiative decays of some of the 
upper levels of these resonantly absorbed K lines.  As pointed out by \cite{lie03}, this may limit 
or negate the effectiveness of Auger destruction.  In addition, the presence of alternate decays 
may increase the total K line flux from a given ion, since it provides a channel for the escape of 
line photons which may be less optically thick than the main channel.  

This effect is displayed in figure~9, which shows the dependence of the line formation 
efficiency on line optical depth.  This figure was calculated by taking a model slab 
calculated with {\sc xstar}, 
initially chosen to have log($\xi$)=2. and column density 10$^{21}$ cm$^{-2}$, and progressively 
increasing the optical depths of all the lines by a multiplicative factor which 
is displayed as the horizontal axis of the figure.  This shows that, for the first row ions which 
dominate the ionization distribution under these conditions, Auger destruction does result in a 
net suppression of most lines over the range of multiplication factors shown here.   However,
the effect requires large enhancements in the optical depth beyond what is found in 
this fiducial slab model.  The reason for this is that the K line optical depth for the strongest lines 
of each ion in the slab is 
$\tau = {{\pi e^2}\over{m_e c}} {{f}\over{\Delta\nu}} N x_l y_{Fe} \simeq 0.08 N_{21}$, 
for a line width $\Delta\nu\simeq 10^{14}$ s$^{-1}$ (comparable to the natural radiative width for 
strong K lines and to the Doppler width at 10$^5$K),
$f=0.1$, $x_l=0.1$, $y_{Fe}=3 \times 10^{-5}$, and $N_{21}=N/10^{21}$.  But this neglects the 
fact that most of the ground terms have multiple levels, and hence at least two K line components. 
Increasing the optical depth has the effect of shifting the emitted line  energy from the dominant 
component into the weaker one, such that the sum remains approximately constant.   Auger destruction 
does not become important until the optical depths in both components are greater than unity.  Since 
the fiducial slab has very small population in the lower levels of many of the components, very large
enhancements in the optical depths are needed in order to suppress the net emitted K line flux.

A more realistic estimate of the importance of Auger destruction comes from the results of 
a family of model slabs calculated as a function of column 
density.  Results of such a calculation are shown in figure~10, which displays the line 
emissivity per ion summed over the lines of various ions.  This figure shows that at 
the largest columns there is a net decrease in the 
line emissivity, particularly for the first row.  But in this case the effect is primarily due to the 
change in the ionization balance of the slab; at high columns the average ionization of the slab 
decreases owing to the shielding of the deeper regions.  Evidence for this comes from the fact that the 
K lines due to the second and third row ions increase at large columns, as the first row ion lines 
decrease.  The total emissivity of the slab, summed over all ions, does decrease at the highest 
ionization parameters, by a factor of order unity.
This suggests that ionization balance effects are more important 
than Auger destruction in determining the strengths of the K lines from first row ions in 
real slabs.  However, this conclusion depends on the details of the conditions in the emitting gas.
If the temperature and density were greater than what we consider, then the populations of all the 
K line lower levels can be comparable.  If so, Auger destruction may have a comparable 
effect on all the K lines of a given ion.

\subsection{Model Emission Spectra}

Figure 11 shows model emission spectra calculated for  model slabs  with 
various ionization parameters in the range -2 $\leq$ log($\xi$) $\leq$ 3.    All 
the slabs are geometrically thin, in the sense that the physical 
thickness of the slab is small compared with the distance from the continuum source.  
In this case we plot the specific line emissivity, $j_{\varepsilon}/n^2$ in 
erg cm$^3$ s$^{-1}$ erg$^{-1}$ as a function of energy.  
Slabs with ionization parameters log($\xi$)$\leq$1 have ionization balance dominated by the third 
row ions and display K line spectra consisting of the K$\alpha_1$ and K$\alpha_2$ lines.  Second row ions, 
important at ionization parameters greater than log($\xi$)=1, cause a blue wing on the line and 
narrow emission components at 6.45 - 6.5 keV.  First row ions produce lines from 6.5 - 7 keV.

For log($\xi$)=-2 the ionization balance is dominated by Fe III and Fe II.  The K$\alpha$ lines from 
these two ions are apparent in the figure,  K$\alpha_1$ at 6.406 keV (1.936 $\AA$) from Fe III and at 
6.404 keV (1.936 $\AA$) from Fe IV are indistinguishable at this scale and are of comparable intensity. 
The K$\alpha_2$ at 6.39 keV is dominated by Fe IV (1.940 $\AA$).  
The ratio K$\alpha_2$/K$\alpha_1$ is 0.65 for Fe IV, and 0.40 for Fe III.
The K$\beta$ complex is dominated by the 7.061 keV (1.755 $\AA$)  K$\beta_1$
line of both ions, with smaller contributions from K$\beta_2$ at 7.055 keV (1.757 $\AA$).  At this 
ionization parameter there are also contributions from Fe IV, which has line energies which are 
indistinguishable from those of lower ionization stages at this resolution.  Also apparent in this 
panel is the set of lines at 6.402 keV (1.937 $\AA$) from Fe V, which are distinct from the 
complex produced by lower ion stages.  Fe V also has component of K$\alpha_2$ which appear at 6.391 keV
(1.940 $\AA$), producing a  high energy component of this complex, and other components near 
6.387 keV, which is lower than the energy of the Fe II-IV K$\alpha_2$.   For Fe V the ratio 
K$\alpha_2$/K$\alpha_1$ is 0.53.  
The K$\beta$ line of this ion makes a high energy shoulder on the complex at 7.066 KeV (1.755 $\AA$).
The profiles shown in figure 11 are scaled in the vertical direction for ease in viewing, so the 
panels have differing vertical axis scales.  These reflect the approximate factor of 2-3  range in the 
line emissivity for various ions (as shown in figure 7) and also the differing spread in energy 
of the emissivity across the panels.  

For log($\xi$)=-1.55, the ionization balance is dominated by Fe IV and by Fe V.  As a result the components 
due to Fe V, which are discernibly different from Fe IV and below, become more apparent than 
at lower ionization parameters.  This includes the 6.402 keV K$\alpha_1$ component, the 
6.387 keV K$\alpha_2$ component, and the 7.04 and 7.061 keV K$\beta$ components.  
For log($\xi$)=-1.1 the ionization balance is dominated by Fe V and  VI, and the line centroids 
remain almost unchanged.  The lines are all slightly narrower in these panels, owing to the reduced 
blending of line components of different energy.
For log($\xi$)=-0.65 The contribution of Fe IX results in K$\alpha_1$ components near 6.398 keV
and  Fe VIII produces distinct K$\alpha_2$ components at 6.387 keV.  Fe VIII also produces a broader 
distribution of narrow K$\alpha_2$ components than lower ion stages.  K$\beta$ components 
at 7.081 and 7.086 keV are present in addition to those produced by Fe V, VI.
At log($\xi$)=-0.2 the K$\alpha$ line is dominated by Fe IX at 6.396 and 6.388 keV.  The 
centroid has the lowest energy of the models shown in figure 11, and is broader than at lower 
ionization parameters due to blending with higher energy components from Fe XIII and Fe X.
The K$\beta$ line shows components at higher energy than in the previous panels, due to 
emission from Fe IX.  Thus Fe IX simultaneously has lower energy K$\alpha$, and higher energy 
K$\beta$ than the lower ion stages of the third row.
Fe IX and X have K$\alpha$ energies which are indistinguishable and they continue to dominate 
the line emission at log($\xi$)=0.25.  The result is that the K$\alpha$ complex changes little 
through this range of ionization parameter.  The K$\beta$ increases in energy from 7.085
to 7.098 (Fe XI)  and 7.110 keV (Fe XII).  At log($\xi$)=0.2 the strongest K$\alpha$ component is 
still Fe X at 6.388 keV, but there is a significant contribution 
from higher energies: Fe XII at 6.408,  Fe XIII at 6.411 keV, Fe XIV at 6.416 keV.
Throughout the region of parameter space where third row ions dominate, the total width of the 
K$\alpha$ blend is $\sim$20 eV.  This corresponds to a Doppler velocity (FWHM) of $\sim$900 km s$^{-1}$, and 
represents the minimum width that can be expected from fluorescence by low ionization iron.  

At ionization parameters where the second row ions dominate the ionization balance, the 
character of the line profile undergoes a qualitative change.  The relatively simple 
K$\alpha_1$/K$\alpha_2$
shape is replaced by a blend of narrow lines which is generally broader, which lacks a regular 
universal shape, but where blending often makes identification of all the lines 
impossible at the resolution we display.  The onset of this behavior begins at log($\xi$)=0.7, where  lines from 
second row ions are comparable in prominence with lines from third row ions.  
At log($\xi$)=1.15 -- 1.45 second row ions dominate, with varying intensity ratios:
Fe XVI at 6.426 keV (1.929 $\AA$), Fe XVII 6.430 keV (1.928 $\AA$), Fe XVI 6.414 keV (1.933 $\AA$), 
and Fe XVIII 6.435 keV (1.927 $\AA$).  The K$\beta$ spectrum is nearly constant in this range of 
ionization parameter, and is dominated by Fe XV 7.158 keV (1.732$\AA$).

When the first row ions dominate, it is generally possible to descriminate individual line components, 
and the spread in energy is greater than at lower ionization parameter.
For log($\xi$)=1.75 and greater the ionization balance is dominated by first row ions: Fe XIX 6.466, 
6.47 keV (1.916, 1.918 $\AA$).  At log($\xi$)=2.2 the Fe XX complex near 6.5 keV is apparent, and 
at log($\xi$)=2.35 and above lines due to Fe XXI near 6.55 keV (1.894 $\AA$), Fe XXII near 6.58 keV
(1.882 $\AA$), and Fe XXIII near 6.63 keV (1.870 $\AA$).  At ionization parameters log($\xi$)$\geq$2.8 the 
emission is dominated by the  lines of Fe XXV near 6.7 keV.

To summarize the results of this section:  The behavior of the line profile can 
be crudely divided into 3 ionization parameter ranges, depending on which ions dominate 
the ionization balance.  These are distinct in their character, and thus provide potential 
diagnostics of the ionization parameter in observed spectra.  
At low ionization parameter, log($\xi$)$\leq$1, the line profile has a characteristic shape 
consisting of the K$\alpha_1$ and K$\alpha_2$ components.  The separation of these two 
components is approximately constant at $\simeq$13 -- 15 eV, and the ratio varies slowly 
with ion stage, from 2:1 for Fe II to 1.7:1 for Fe IX.  The energies of the K$\alpha$ components moves 
down with increasing ionization, by $\simeq$5 eV, across the third row, due to the interaction between the 
2p and the 3d orbitals.  The K$\beta$ energy increases monotonically with increasing ionization.  
The character of the profile changes qualitatively at log($\xi$)$\geq$1,  where the second row ions 
cause the line centroid to move 
up in energy by a significant amount.  The line shape consists of a blend of narrow 
components with total width 20-30 eV, which may still appear as a single line if observed by an instrument with 
moderate spectral resolution or statistics. If represented by a single broad line component, these blends 
would have an apparent width $\sim$1000 -- 1500 km s$^{-1}$.  The profile changes 
qualitatively again at log($\xi$)$\geq$2, where the first row ions 
dominate.  Here distinct contributions to the line from the various ionization stages are separated in 
energy by $\sim$ 50 -- 75 eV, and are less likely to be interpreted as part of a single feature.

\subsection{Model Absorption Spectra}

Figure 12 shows the results of model slab calculations for absorption.  These are calculated from 
a family of model slabs with varying ionization parameter and a column density of 10$^{23.5}$ cm$^{-2}$.
Since these model slabs are all optically thick, they do not contain gas with a constant ionization balance 
throughout.  In this sense the results are different from those shown in the previous subsection for 
emission.  As a result, features due to low ionization iron are present in the spectra even when the 
ionization parameter is greater than the value where these ions would occur in an optically thin 
situation.  

The absorption spectra are dominated by resonance structure due to 1s-np auto ionizing transitions.
Few instances of detectable edges are seen, in the sense of sharp step functions in the opacity 
above a threshold energy.  At low ionization parameter, where the third row ions dominate, the 
resonances are due to 1s-4p transitions.  At the lowest ionization, the edges of adjacent ion stages
have threshold energies separated by $\sim$20-30 eV.  The ionization balance results in a mix of 
3-5 adjacent ionization stages at a given ionization parameter, so the edge is broadened to 
$\sim$50-60 eV.    As ionization parameter increases above log($\xi$)=-2 the edge and 
resonance structure gradually 
broadens above 7.14 keV, the resonances become more prominent and more widely spaced.
The K$\alpha$ line is absent, since these ions have closed 2p sub shells.
As the ionization parameter increases in progressing panels of figure~12, the 
depth of the resonance feature near 7.1 keV increases, and the resonance structure above 7.2 keV 
becomes more pronounced.  Owing to the absorption by nearly neutral iron in the shielded parts 
of the clouds, the features due to near-neutral Fe II and Fe III remain prominent for log($\xi$)$<$1.   

For log($\xi$)$>$1 features from second row ions become prominent, notably K$\alpha$ absorption, along with 
more resonance structure above 7.2 keV, and for log($\xi$)$>$2 the edge structure is dominated 
by resonances spread over $\Delta\varepsilon\simeq 500$eV and the sharp neutral edge is not discernible.
At log($\xi$)=1.5 the ionization is dominated by 2nd row ions, Fe XVI, XV, XIV, and the structure 
looks like a K edge near 7.2 keV, with structure due to range of ionization states, plus K$\beta$ below
near 7.1 keV.  At log($\xi$)=1.75 there is structure in both the K edge and K$\beta$ K$\alpha$ is 
apparent near 6.5 keV, due to to a small admixture of first row ions.
For log($\xi$)=2.0 the ionization balance is dominated by Fe XVIII and neighboring ions.  Therefore, there are 
resonances at 7.15, 7.2, 7.3, 7.42, 7.52, 7.65, 7.73, corresponding to 
K$\beta$ from Fe XVII, XVIII, XIX, and 1s-4p for 
Fe XVIII and XIX, respectively.  Notice also the structure in the K$\alpha$ line.
For log($\xi$)=2.5, the ionization balance is dominated by Fe XX and neighboring ions.  Therefore the 
spectrum shows absorption near 7.55 keV, the Fe XXI K$\beta$, plus the other strong resonances 
of this ion near 7.85 keV.  In addition, there are features at 7.60 KeV due to Fe XXII K$\beta$, 
7.75 KeV due to  Fe XXIII K$\beta$.  At log($\xi$)=3 the Fe XXV K edge is present at 8.8 keV, and the 
lower energy features are due to Fe XXIII and Fe XXII.  The Fe XXVI L$\alpha$ line is apparent at 6.97 keV.

In analogy with the previous  section,  the behavior of the opacity can 
be crudely divided into distinct ionization parameter ranges
which are diagnostic of the ionization parameter in observed spectra.  
At low ionization parameter, log($\xi$)$\leq$1, the resonances are close to 
threshold.  The K edge opacity has the character of an edge smeared by $\sim$10 -- 30 eV.
The character of the opacity changes qualitatively at log($\xi$)$\geq$1,  where the second row ions 
cause the edge to be smeared by $\sim$300 -- 500 eV.  K $\beta$ appears in absorption in these ions.
The distribution changes qualitatively again at log($\xi$)$\geq$2, where the first row ions 
dominate.  Here the edge is smeared by $\geq$500 eV, and K$\alpha$ line appears in absorption.
At very high ionization parameters, which we do not model in detail here (log($\xi$)$\geq$3), 
the opacity is dominated by ions which have
a comparatively simple structure, H- and He-like Fe, and the opacity is described by Lyman-series 
absorption plus a few sharp K edges.

\subsection{Simulated Observations}

Many of the features that appear in the model spectra presented so far are on an energy 
scale which is finer than can be resolved with instruments which use, eg. CCD detectors.  This 
is illustrated in Figure 13a, in which we display a simulated observation using the PN 
detector on the XMM satellite, for an assumed source flux of 1.5 $\times 10^{-9}$ erg cm$^{-2}$
s$^{-1}$ and observing time of 10$^5$ seconds.  The absorber is assumed to be the same as shown 
in figure 10, i.e. log($\xi$)=2, N=10$^{23}$ cm$^{-2}$.  This shows that the 
K$\alpha$ line is not resolved, but appears as a single broad feature.   The K edge is 
not readily apparent to the eye, but does cause a significant deviation from the continuum 
power law.   This cannot be distinguished from a superposition of sharp edges corresponding to
the first row ions.   We show in figure 13b what the XRS on Astro-E
would observe with the same source spectrum, flux, and observing time.  
Much of the complexity of the models 
in figure 12 is apparent.  In order to underscore this, we show in Figure 13c 
what the XRS would observe if the spectrum 
were simply a superposition of sharp edges, assuming the same 
flux, observing time, etc.  This model is what 
was predicted by early versions of {\sc xstar} (version 2.1d and before), and the simulated spectrum
is similar to that observed by the XMM PN in panel a. 
using the new data and models presented in the previous subsection
This demonstrates that many features will be detectable
using such an observation.  The distinction between the new data and that which was previously
in use is readily apparent, and the level of complexity which is accessible can constrain the 
ionization of the gas, for example, and likely other properties such as gas flows, relativistic 
effects,  and multiple components.

\section{Summary}

We have calculated the efficiency of iron K line emission and iron K absorption in photoionized 
models using a new and comprehensive set of atomic data.  We have shown that:

\noindent $\bullet$ The average fluorescence 
yield for each ion in the first row is sensitive to the level population distribution produced by 
photoionization, and these yields are generally smaller than those predicted assuming 
the population is according to statistical weight.   

\noindent $\bullet$ The presence of multiple levels in the ground term can lead to density 
dependence of the ratios of various K emission line components from a given ion for densities 
greater than $\sim 10^{12}$ cm$^{-3}$.  This is due to the effect of electron 
collisions in mixing the population  into levels with differing fluorescence yields.  

\noindent $\bullet$ The presence of multiple line components of disparate optical 
depths reduces the influence of 
optical depth because it allows energy to shift from one line component to another within an 
ion as optical depth increases, while maintaining the total emitted line energy approximately 
constant.  

\noindent $\bullet$ The K line spectrum in emission has a high
degree of complexity, particularly for first and second row ions.  Third row ions show a 
shift in the centroid of the K$\alpha$ feature with ionization stage, in the sense that the 
centroid shifts to lower energies as the ion stage increases from neutral, and then back to higher 
energies in the second row.  This can be used as 
an observational diagnostic for relatively bright objects viewed with high resolution 
instruments.  

\noindent $\bullet$ The K shell opacity is dominated by the strongly damped resonances 
at energies below the K ionization threshold.  These conspire to smear and weaken the change in 
opacity at the edge, and also present numerous narrow features in the absorption spectrum.

All of the atomic data used in this calculation is available on line through 
electronic tables associated with the earlier papers in this series, and as part of the topbase 
electronic database (http://heasarc.gsfc.nasa.gov/topbase/topbase.html).  
The results shown in this paper are incorporated into the publicly available 
photoionization and spectrum synthesis code xstar and the associated xspec tables
(http://heasarc.gsfc.nasa.gov/docs/software/xstar/xstar.html).

\begin{acknowledgements}
This work was supported by  grant NRA-00-01-ATP-025 through the NASA Astrophysics Theory Program.
\end{acknowledgements}

\appendix

\section{Appendix: Description of Level Scheme and Photoabsorption features}

Since each ion has, in principle, an infinity of K lines associated with the excited levels 
of the valence shell in addition to transitions to levels with L and M shell vacancies, 
it is necessary to limit the selection of levels and lines in order to make a calculation
of the line spectrum feasible.  In doing so, we assume that all the K 
vacancy levels are populated by photoionization from the next lower ion stage.  Then it is only
necessary to include K vacancy levels which satisfy the selection rules of \cite{rau76}, namely
$\Delta$J=$^+_-$1/2,$^+_-$3/2 and $\Delta$S=$^+_-$1/2.  The valence shell excitation levels are the same as those 
included in the standard {\sc xstar} code, namely the levels of the ground term plus as 
many of the levels with the same principle quantum number as the ground, plus the next greater, 
as is feasible.  The L- and M-vacancy levels are chosen in order to include all of the 
strongest K lines.  A detailed list of the numbers of these levels and the associated 
edge energies  follows.  

Fe ~{\sc ii} has 51 metastable levels grouped within 4 eV of the ground level from which
K-vacancy levels of Fe~{\sc iii} can be populated by photoionization.  
We consider 53 K lines,  7 3p-vacancy levels, 26 L-vacancy levels, 3 K-vacancy levels.  
The absorption spectrum 
is dominated by a sharp K edge at 7.136 keV, although  there is one strong resonance 
very close to the K threshold, corresponding to the 1s-4p unresolved transition array (UTA).
Fe ~{\sc iii}  has 23 metastable levels within 5 eV of the ground level from which K-vacancy levels
can be populated by photoionization.  
 We consider 63 K lines, 13 3p-vacancy levels, 31 L-vacancy levels, 3 K-vacancy levels.
The absorption spectrum is dominated by the 1s-4p resonance below the  K edge at 7.14 keV.
Fe~{\sc iv} has  32 metastable levels within 10 eV of the ground level from which K-vacancy levels
can be populated by  photoionization. 
 We consider 50 K lines, 10 3p-vacancy levels, 22 L-vacancy levels, 3 K-vacancy levels.
 The absorption spectrum is dominated by the 1s-4p,5p UTAs below the  K edge at 7.158, 7.160  keV.
For Fe~{\sc v} we consider 35 K lines, 8 3p-vacancy levels, 20 L-vacancy levels, 2 K-vacancy levels.
 The absorption spectrum is dominated by 1s-$n$p resonances between 7.17-7.19  keV.
Fe~{\sc vi}  has  18 metastable levels within 9 eV of the ground level from which K-vacancy states
can be populated by  photoionization. 
 We consider 38 K lines, 11 3p-vacancy levels, 21 L-vacancy levels, 2 K-vacancy levels.
 The absorption spectrum is dominated by 1s-$n$p resonances between 7.19-7.21  keV.
Fe~{\sc vii}  has  8 metastable levels within 9 eV of the ground level from which K-vacancy levels
can be populated by  photoionization. 
 We consider 57 K lines, 13 3p-vacancy levels, 27 L-vacancy levels, 3 K-vacancy levels.
 The absorption spectrum is dominated by 1s-$n$p ($n$=4--8) resonances between 7.21-7.24  keV.
Fe~{\sc viii} (Ca-like) has  1 metastable level within 1 eV of the ground level from which 
K-vacancy levels can be populated by  photoionization. 
 We consider 80 K lines, 11 3p-vacancy levels, 25 L-vacancy levels, 3 K-vacancy levels.
 The absorption spectrum is dominated by 1s-$n$p ($n$=4--7) resonances between 7.232-7.28  keV.
Fe~{\sc ix} (K-like) has no metastable state within 10 eV of the ground level. 
We consider 35 K lines, 12 3p-vacancy levels, 10 L-vacancy levels, 3 K-vacancy levels. 
The absorption spectrum is dominated by 1s-$n$p ($n$=4--7) resonances between 7.232-7.28  keV.

Fe~{\sc x} is a second row ion which has a ground configuration characterized by 
an open 3p valence shell. The ground term is a doublet split by 1.95 eV. There
are 4 K lines that connect the 2 ground-term levels and 2 L-vacancy levels to one K-vacancy level.
The absorption spectrum
is dominated by the K$\beta$ UTA at 7.089 keV, plus 4 distinct resonances
between 7.25-7.40  keV. 
The ground configuration in Fe~{\sc xi} (S-like) has 5 fine-structure levels split by 10 eV.
There are 44 K lines connecting the above-mentioned 5 levels and 10 L-vacancy levels to 4 K-vacancy levels. 
The absorption spectrum
is dominated by the K$\beta$ UTA at 7.1 keV, plus  resonances
between 7.3-7.45  keV.
In Fe~{\sc xii} (P-like), the ground configuration has 5 levels split by 10 eV. There
are 107 K lines connecting the above-mentioned levels and 21 L-vacancy levels to 5 K-vacancy levels.  
The absorption spectrum
is dominated by the K$\beta$ UTA at 7.12 keV, plus resonances
between 7.4-7.5  keV.
The ground configuration in Fe~{\sc xiii} (Si-like)  has 5 levels split by 11 eV. There
are 53 K lines connecting the above-mentioned levels and 27 L-vacancy levels to 2 K-vacancy levels.
The absorption spectrum
is dominated by the K$\beta$ UTA at 7.12 keV, plus resonances
between 7.35-7.5  keV.
In Fe~{\sc xiv} (Al-like), the ground configuration has only one doublet split by 2 eV. There
are 90 K lines connecting the ground-term levels and 21 L-vacancy levels to 5 K-vacancy levels.
The absorption spectrum
is dominated by the K$\beta$ UTA at 7.14 keV, plus  resonances
between 7.35-7.55  keV.
The ground configuration in Fe~{\sc xv} (Mg-like) is characterized by closed shells and therefore has only
a $^1{\rm S}_0$ level. There are 56 K lines that connect the ground level, the 4 levels belonging
to the excited $3{\rm p}^2$ valence configuration and 10 L-vacancy levels to 4 K-vacancy levels.  
The absorption spectrum is dominated by the K$\beta$ UTA at 7.15 keV,
 plus  resonances between 7.4-7.6  keV.
In Fe~{\sc xvi} (Na-like),  the ground configuration, i.e. 3s,  has only a $^2{\rm S}_{1/2}$ level. There
are 4 K lines connecting the 3p~$^2{\rm P}$ levels and 2 L-vacancy levels to one K-vacancy level.  
The absorption spectrum
is dominated by one K$\beta$ line at 7.14 keV, plus 4 distinct resonances
between 7.4-7.6  keV.
Fe~{\sc xvii} (Ne-like)  has a ground configuration characterized by closed shells
which has therefore one $^1{\rm S}_0$ level. The K-vacancy states correspond to the coupling 
of a [1s] hole with a M-shell electron.  The L-vacancy levels are located above 0.7 
keV from the ground level. It has 38 K lines that connect the ground level, 
the 10 levels belonging to the excited [2p]3p configuration and 26 L-vacancy levels to 6 K-vacancy levels.
Absorption spectrum
is dominated by the K$\beta$ UTA at 7.19 keV, plus 4 distinct resonances
between 7.4-7.55 keV.

Fe~{\sc xviii} (F-like) is a first row ion which is characterized by an open L shell. 
The ground configuration has one hole in the 2p sub-shell and therefore has only one
term which is a $^2{\rm P}$ split by 13 eV. There are 2 K lines connecting the ground term
to one K-vacancy level. The absorption spectrum is dominated by the K$\alpha$ lines at 6.42 and
6.43 keV and by the K$\beta$ resonances at 7.27 keV, plus distinct resonances of higher 
members of the 1s-$n$p series at 7.55, 7.65 and 7.75  keV.
In Fe~{\sc xix} (O-like), the ground configuration has 5 fine-structure levels split by 40 eV.
There are 40 K lines connecting the 10 levels of the $1{\rm s}^2({\rm 2s2p})^6$ Layzer complex
to 6 K-vacancy levels. The absorption spectrum is dominated by the K$\alpha$ transition array and
by the K$\beta$ UTA at 7.40 keV, plus  resonances between 7.65--8.0 keV.
The ground configuration in Fe~{\sc xx} (N-like) has  5 levels split by 40 eV.
There are 101 K lines that connect the 15 levels of the $1{\rm s}^2({\rm 2s2p})^5$ Layzer complex
to 16 K-vacancy levels. The absorption spectrum is dominated by the K$\alpha$ transition array and
by the K$\beta$ UTA at 7.45 keV, plus resonances between 7.75--8.1 keV.
In Fe~{\sc xxi} (C-like), the ground configuration has  5 fine-structure levels split by 46 eV.
There are 218 K lines connecting the 20 levels of the $1{\rm s}^2({\rm 2s2p})^4$ Layzer complex
to 30 K-vacancy levels. The absorption spectrum is dominated by the K$\alpha$ transition array and
by the K$\beta$ UTA at 7.55 keV, plus resonances between 7.8--8.3 keV.
In Fe~{\sc xxii} (B-like), the ground configuration has  one LS term which is a $^2{\rm P}$ split by
15 eV. There are  218 K lines that connect the 15 fine-structure levels of 
the $1{\rm s}^2({\rm 2s2p})^3$ Layzer complex to 34 K-vacancy levels.
The absorption spectrum is dominated by the K$\alpha$ transition array and
by the K$\beta$ UTA at 7.65 keV, plus  resonances between 7.9--8.4 keV.
The ground configuration in Fe~{\sc xxiii} (Be-like) is characterized by closed shells and therefore
has one $^1{\rm S}_0$ level. There are 101 K lines connecting the 10 levels of 
the $1{\rm s}^2({\rm 2s2p})^2$ Layzer complex to 30 K-vacancy levels.
The absorption spectrum is dominated by the K$\alpha$ transition array and
by the K$\beta$ UTA at 7.75 keV, plus resonances between 8.1--8.55 keV.
In Fe~{\sc xxiv} (Li-like), the ground configuration, 2s, has one level which is a $^2{\rm S}_{1/2}$.
There are 25 K lines that connect the 3 fine-structure levels of
the $1{\rm s}^2({\rm 2s2p})^1$ Layzer complex to 16 K-vacancy levels.
For this ion the resonances are damped only by radiative damping, and so are relatively unimportant compared 
with the K photoionization edge.


\clearpage
\begin{deluxetable}{lrrr}
\tablecaption{\label{feat} Strongest near K-edge absorption features in the iron ions
that can be resolved with a resolving power of 1000. The
line parameters, i.e. the wavelength, oscillator strength and full width at half maximum,  are fitted values extracted from gaussian profiles using RAP cross sections with $\Delta~=~0.001$.}
\tablewidth{0pt}
\tablehead{ \colhead{Ion}&\colhead{$\lambda$}&\colhead{$f$}&\colhead{$\Gamma$}\\
\colhead{}&\colhead{(\AA)}& & \colhead{(m\AA)}
} \startdata
Fe~{\sc ii}    & 1.738 & 0.0041 & 0.021\\
Fe~{\sc iii}   & 1.736  & 0.0050 & 0.021\\
Fe~{\sc iv}    & 1.733 & 0.0060 & 0.021\\
Fe~{\sc v}     & 1.728 & 0.0081 & 0.021\\
Fe~{\sc vi}    & 1.724 & 0.0103 & 0.021\\
Fe~{\sc vii}   & 1.719 & 0.0125 & 0.021\\
Fe~{\sc viii}  & 1.714 & 0.0148 & 0.021\\
Fe~{\sc ix}    & 1.718 & 0.0286 & 0.271\\
                       & 1.694 & 0.0146 & 0.296\\
Fe~{\sc x}     & 1.712 & 0.0279 & 0.270\\
                       & 1.694 & 0.0146 & 0.287\\
                       & 1.688 & 0.0130 & 0.340\\
Fe~{\sc xi}    & 1.707& 0.0133 & 0.315\\
                      & 1.693 & 0.0068 & 0.337\\
                      & 1.686 & 0.0134 & 0.321\\
                      & 1.680 & 0.0300 & 0.537\\
Fe~{\sc xii}   & 1.685 & 0.0139 & 0.288\\
                       & 1.670 & 0.0128 & 0.364\\
Fe~{\sc xiii}  & 1.694 & 0.0197 & 0.287\\
                       & 1.677 & 0.0233 & 0.272\\
                       & 1.662 & 0.0309 & 0.758\\
Fe~{\sc xiv}   & 1.688 & 0.0175 & 0.290\\
                        & 1.670 & 0.0328 & 0.265\\
                        & 1.660 & 0.0072 & 0.342\\
                        & 1.651 & 0.0177 & 0.515\\
Fe~{\sc xv}    & 1.681 & 0.0079 & 0.305\\
                        & 1.654 & 0.0050 & 0.353\\
                        & 1.645 & 0.0040 & 0.375\\
                        & 1.633 & 0.0067 & 0.421\\
Fe~{\sc xvi}   & 1.675 & 0.0377 & 0.276\\
                       & 1.654 & 0.0182 & 0.282\\
                       & 1.643 & 0.0131 & 0.318\\
\enddata
\end{deluxetable}

\clearpage
\addtocounter{table}{-1}
\begin{deluxetable}{lrrr}
\tablecaption{\label{feat} Continued.}
\tablewidth{0pt}
\tablehead{ \colhead{Ion}&\colhead{$\lambda$}&\colhead{$f$}&\colhead{$\Gamma$}\\
\colhead{}&\colhead{(\AA)}& & \colhead{(m\AA)}
} \startdata
Fe~{\sc xvii}  & 1.669 & 0.0444 & 0.265\\
                        & 1.647 & 0.0219 & 0.267\\
                        & 1.635 & 0.0133 & 0.285\\
                        & 1.628 & 0.0107 & 0.342\\
Fe~{\sc xviii} & 1.701 & 0.1344 & 0.317\\
                       & 1.644  & 0.0445 & 0.397\\
                       & 1.622 & 0.0236 & 0.467\\
Fe~{\sc xix}   & 1.686 & 0.0052 & 0.226\\
                       & 1.681 & 0.0198 & 0.591\\
                       & 1.625 & 0.0042 & 0.239\\
                       & 1.619 & 0.0058 & 0.527\\
                       & 1.599 & 0.0012 & 0.208\\
                       & 1.594 & 0.0036 & 0.686\\
                       & 1.585 & 0.0026 & 0.377\\
Fe~{\sc xx}    & 1.665 & 0.0583 & 0.318\\
                        & 1.659 & 0.0712 & 0.391\\
                        & 1.602 & 0.0295 & 0.265\\
                        & 1.595 & 0.0249 & 0.377\\
                        & 1.575 & 0.0188 & 0.268\\
                        & 1.567 & 0.0124 & 0.373\\
                        & 1.561 & 0.0086 & 0.322\\
                        & 1.553 & 0.0113 & 0.358\\
Fe~{\sc xxi}   & 1.643 & 0.1283 & 0.466\\
                       & 1.577 & 0.0612 & 0.433\\
                       & 1.547 & 0.0249 & 0.587\\
                       & 1.533 & 0.0171 & 0.353\\
Fe~{\sc xxii}  & 1.627 & 0.1666 & 0.367\\
                       & 1.556 & 0.0521 & 0.414\\
                       & 1.526 & 0.0243 & 0.417\\
                       & 1.510 & 0.0154 & 0.489\\
Fe~{\sc xxiii} & 1.602 & 0.0966 & 0.257\\
                        & 1.531 & 0.0228 & 0.234\\
                        & 1.500 & 0.0042 & 0.276\\
                        & 1.484 & 0.0040 & 0.284\\
\enddata
\end{deluxetable}

\clearpage
\begin{deluxetable}{lrr}
\tablecaption{\label{feat} Average fluorescence yields, weighted by excitation rate.  Yield 1: upper levels are populated by photoionization; Yield 2: upper levels are populated statistically}
\tablewidth{0pt}
\tablehead{ \colhead{Ion}&\colhead{Yield 1}&\colhead{Yield 2}
} \startdata
2&0.34&0.33\\
3&0.34&0.33\\
4&0.34&0.34\\
5&0.35&0.35\\
6&0.35&0.35\\
7&0.33&0.33\\
8&0.35&0.35\\
9&0.36&0.36\\
10&0.36&0.36\\
11&0.36&0.36\\
12&0.36&0.36\\
13&0.37&0.37\\
14&0.37&0.37\\
15&0.37&0.37\\
16&0.38&0.38\\
17&0.39&0.39\\
18&0.39&0.39\\
19&0.48&0.52\\
20&0.41&0.45\\
21&0.31&0.48\\
22&0.31&0.53\\
23&0.26&0.62\\
24&0.13&0.71\\
\enddata
\end{deluxetable}


\clearpage

\begin{figure}[!bp]
\epsscale{0.99}
\plotone{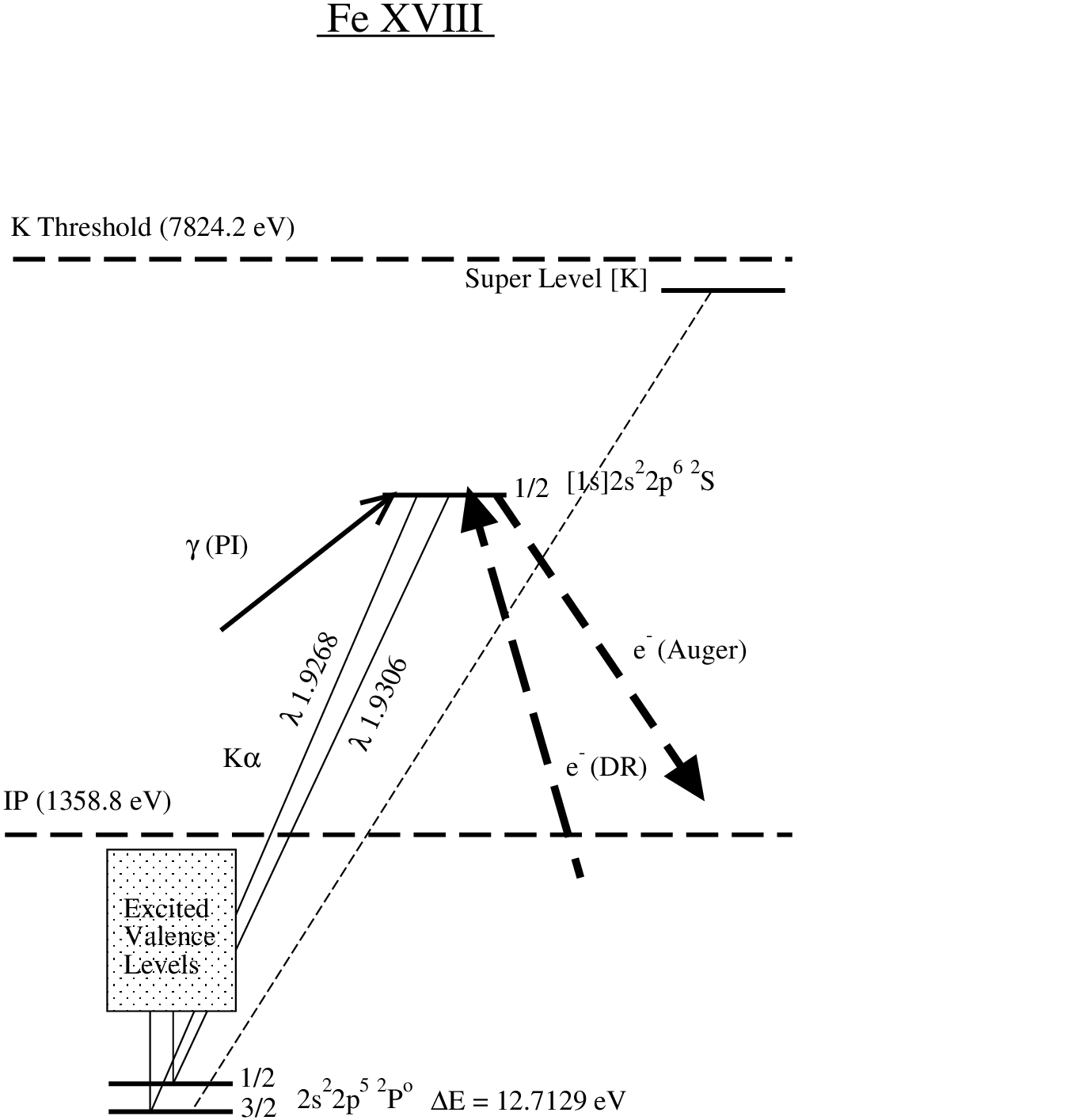}
\caption{Panel A: Simple Grotrian diagram for inner shell processes in Fe~{\sc xviii}}
\end{figure}

\clearpage
\setcounter{figure}{0}

\begin{figure}[!bp]
\epsscale{0.99}
\plotone{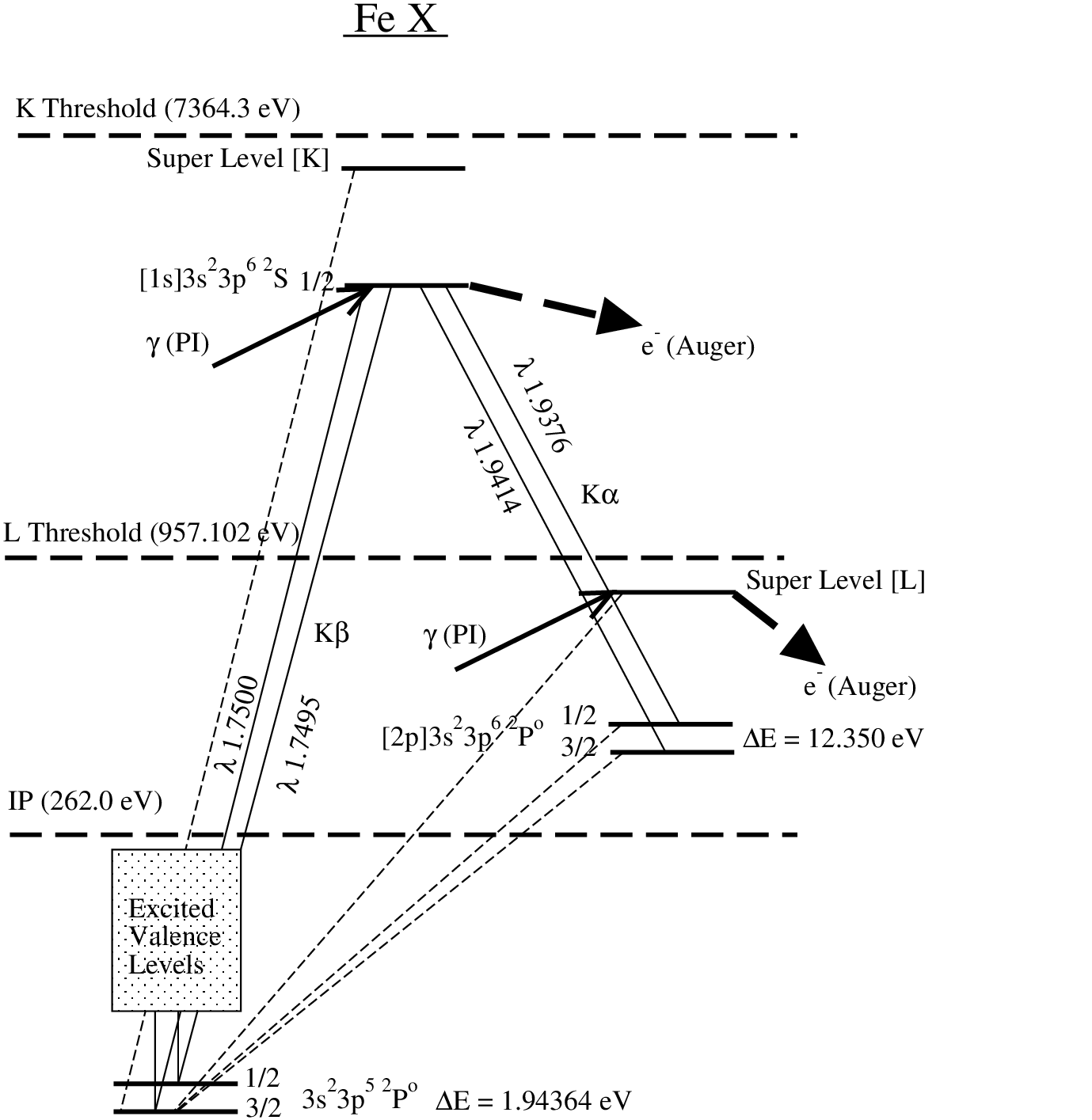}
\caption{Panel B: Simple Grotrian diagram for inner shell processes in Fe~{\sc x}}
\end{figure}

\clearpage
\setcounter{figure}{1}

\begin{figure}[!bp]
\epsscale{0.99}
\plotone{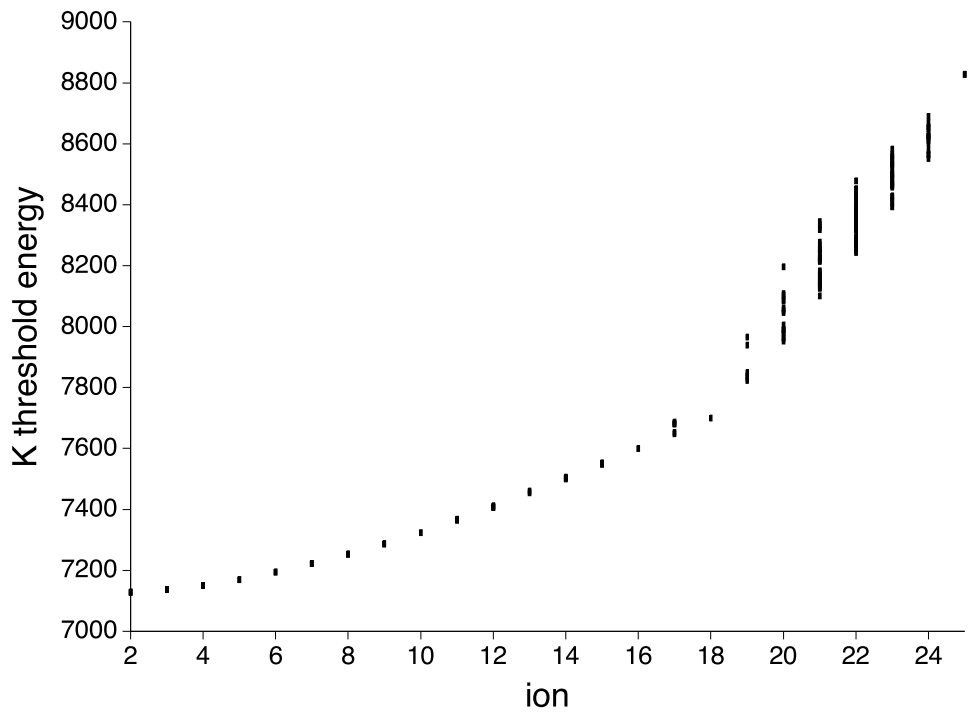}
\caption{K shell ionization thresholds (eV) vs. ionization state.}
\end{figure}

\clearpage

\begin{figure}[!bp]
\epsscale{0.99}
\plotone{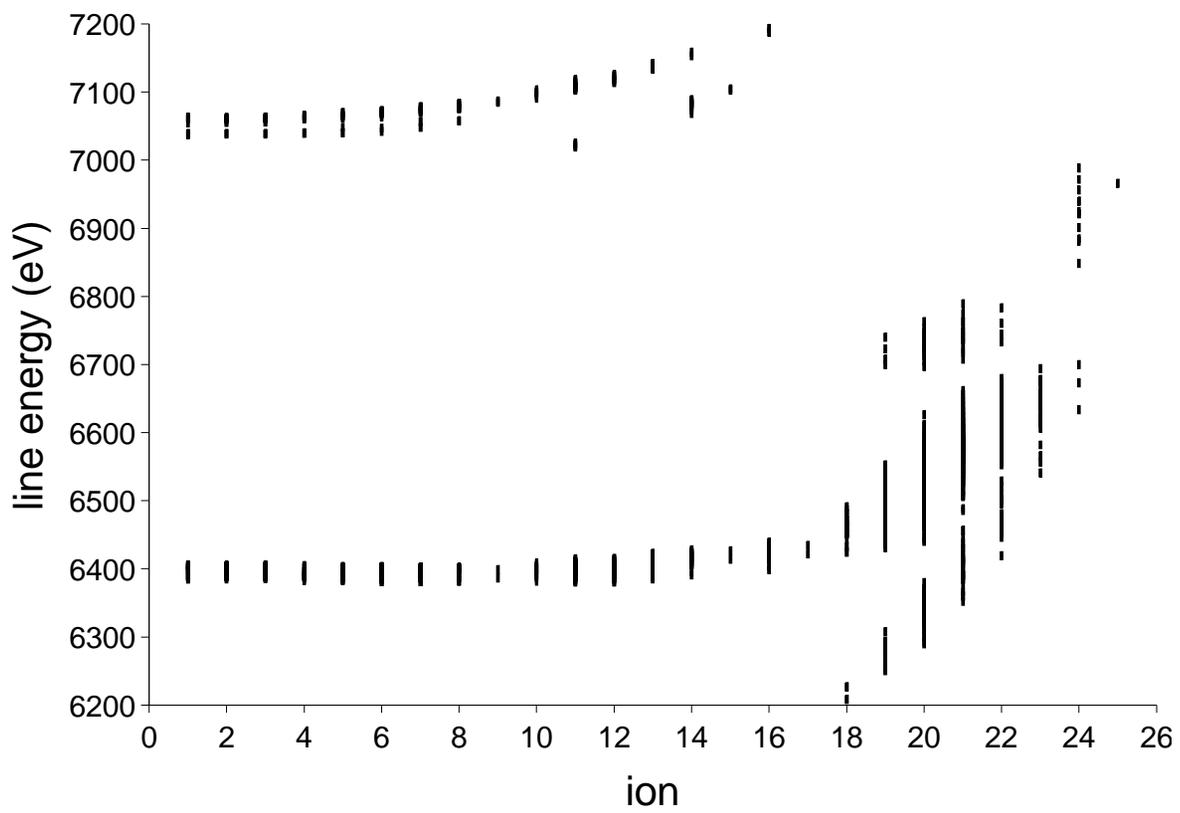}
\caption{K line energy (eV) vs. ionization state.}
\end{figure}

\clearpage

\begin{figure}[!bp]
\epsscale{0.99}
\plotone{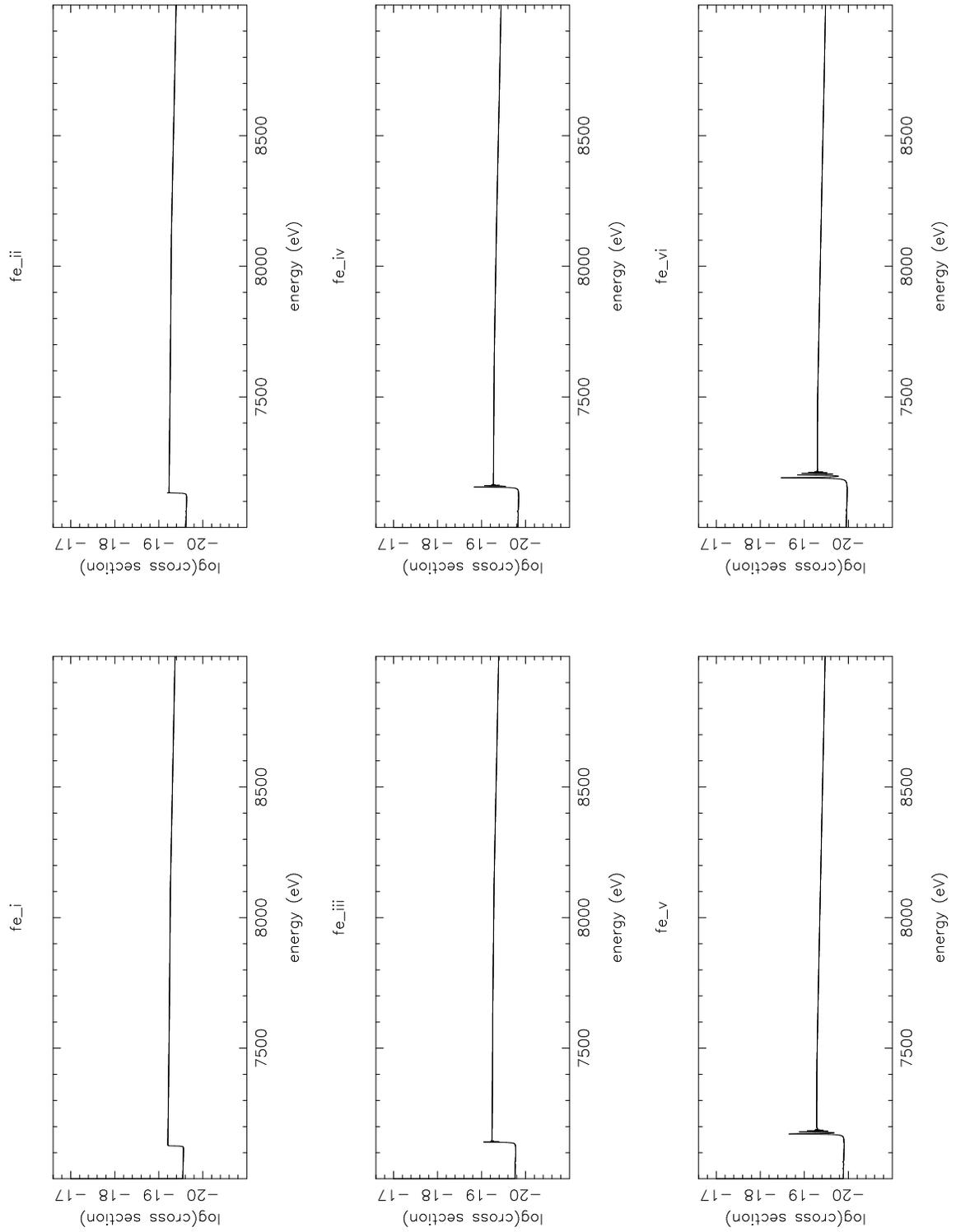}
\caption{Photoionization cross section in the K edge region vs. energy.  Units 
of the cross section are cm$^{-2}$.}
\end{figure}

\clearpage
\setcounter{figure}{3}

\begin{figure}[!bp]
\epsscale{0.99}
\plotone{f4b.epsi}
\caption{continued}
\end{figure}

\clearpage
\setcounter{figure}{3}

\begin{figure}[!bp]
\epsscale{0.99}
\plotone{f4c.epsi}
\caption{continues}
\end{figure}

\clearpage
\setcounter{figure}{3}

\begin{figure}[!bp]
\epsscale{0.99}
\plotone{f4d.epsi}
\caption{continued}
\end{figure}

\clearpage

\begin{figure}[!bp]
\epsscale{0.6}
\plotone{f5.epsi}
\caption{Ion fractions vs. ionization parameter for iron.  Based on 
an $F_\varepsilon \propto \varepsilon^{-1}$ ionizing continuum}
\end{figure}

\clearpage
\begin{figure}[!bp]
\epsscale{0.6}
\plotone{f6.epsi}
\caption{Temperature vs. ionization parameter for iron.  Based on 
an $F_\varepsilon \propto \varepsilon^{-1}$ ionizing continuum}
\end{figure}

\clearpage
\begin{figure}[!bp]
\epsscale{0.6}
\plotone{f7.epsi}
\caption{Line emissivities per ion  vs. ionization parameter for iron.  
 For each ion, 
the emissivity of all K lines summed is plotted.  Emissivity
has been divided by ionization parameter for ease in plotting.}
\end{figure}

\clearpage

\begin{figure}[!bp]
\epsscale{0.6}
\plotone{f8.epsi}
\caption{Dependence of line emissivities on gas density.  For each ion, 
the ratio of the emissivity of all K lines summed to the corresponding quantity 
at density 10$^8$ cm$^{-3}$ is plotted.}
\end{figure}

\clearpage

\begin{figure}[!bp]
\epsscale{0.6}
\plotone{f9.epsi}
\caption{Dependence of line emissivities on optical depth. 
As described in the text, the optical depth of each line calculated for 
a low-column slab is boosted by a depth multiplier.  For each ion, 
the ratio of the emissivity of all K lines summed to the corresponding quantity 
 when the depth multiplier is unity is plotted.}
\end{figure}

\clearpage

\begin{figure}[!bp]
\epsscale{0.6}
\plotone{f10.epsi}
\caption{Dependence of line emissivities on cloud column 
density.  Units are the same as figure 7, averaged over the 
cloud volume.}
\end{figure}

\clearpage
\begin{figure}[!bp]
\epsscale{0.6}
\plotone{f11a.epsi}
\caption{Emission line profiles for a family of 
thin (N=10$^{17}$ cm$^{-2}$) slabs is plotted for various 
ionization parameter.  Vertical scale is arbitrary, but is the 
same for all panels.}
\end{figure}

\clearpage
\setcounter{figure}{10}

\begin{figure}[!bp]
\epsscale{0.6}
\plotone{f11b.epsi}
\caption{continued}
\end{figure}

\clearpage
\setcounter{figure}{10}

\begin{figure}[!bp]
\epsscale{0.6}
\plotone{f11c.epsi}
\caption{continued}
\end{figure}

\clearpage
\setcounter{figure}{10}

\begin{figure}[!bp]
\epsscale{0.6}
\plotone{f11d.epsi}
\caption{continued}
\end{figure}

\clearpage

\begin{figure}[!bp]
\epsscale{0.6}
\plotone{f12a.epsi}
\caption{Absorption spectra for a family of 
thick (N=3 $\times$ 10$^{23}$ cm$^{-2}$) slabs is plotted for various 
ionization parameter.  Vertical scale is arbitrary, but is the 
same for all panels.}
\end{figure}

\clearpage
\setcounter{figure}{11}

\begin{figure}[!bp]
\epsscale{0.6}
\plotone{f12b.epsi}
\caption{continued}
\end{figure}

\clearpage
\setcounter{figure}{11}

\begin{figure}[!bp]
\epsscale{0.6}
\plotone{f12c.epsi}
\caption{continued}
\end{figure}

\clearpage
\begin{figure}[!bp]
\epsscale{0.6}
\plotone{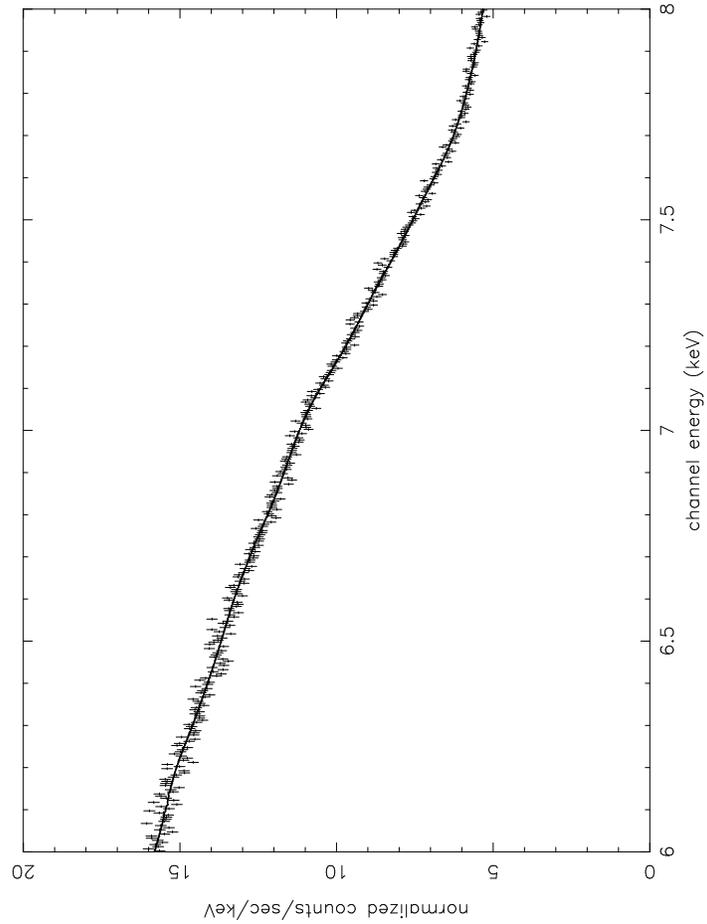}
\caption{Panel A:  Simulated spectra of a 100 mCrab source with a spectrum similar to 
that shown in figure 12, log($\xi$)=2, log(N)=23, as observed with the 
PN instrument on XMM/Newton for 10$^5$ seconds.}
\end{figure}

\clearpage
\setcounter{figure}{12}

\begin{figure}[!bp]
\epsscale{0.6}
\plotone{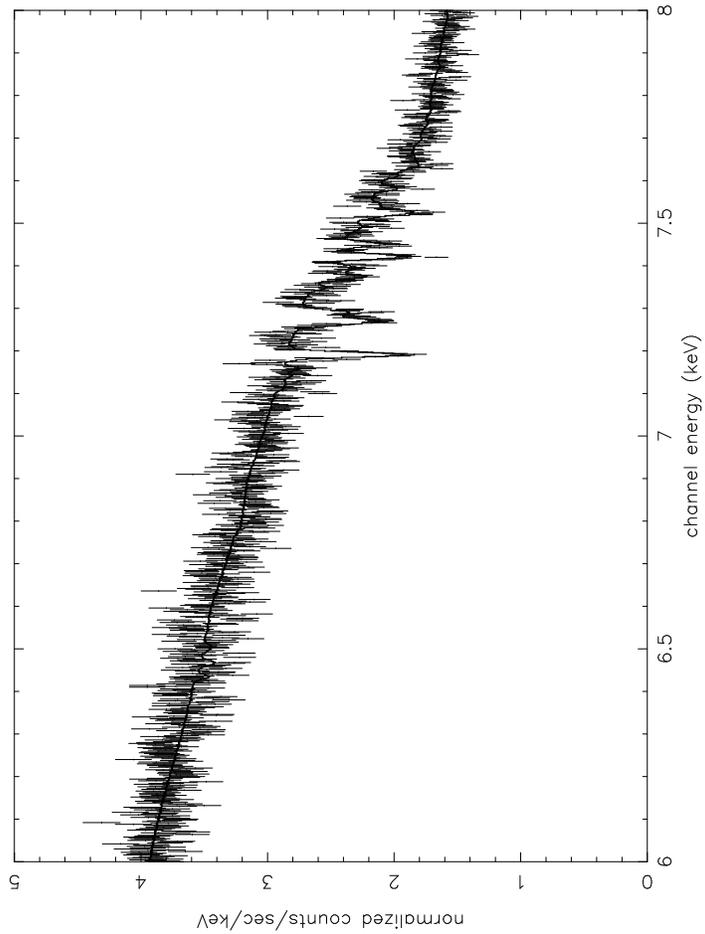}
\caption{Panel B: Simulated spectra of a 100 mCrab source with a spectrum similar to 
that shown in figure 12, log($\xi$)=2, log(N)=23, as observed with the 
XRS instrument on Astro-E2 for 10$^5$ seconds.}
\end{figure}

\clearpage
\setcounter{figure}{12}

\begin{figure}[!bp]
\epsscale{0.6}
\plotone{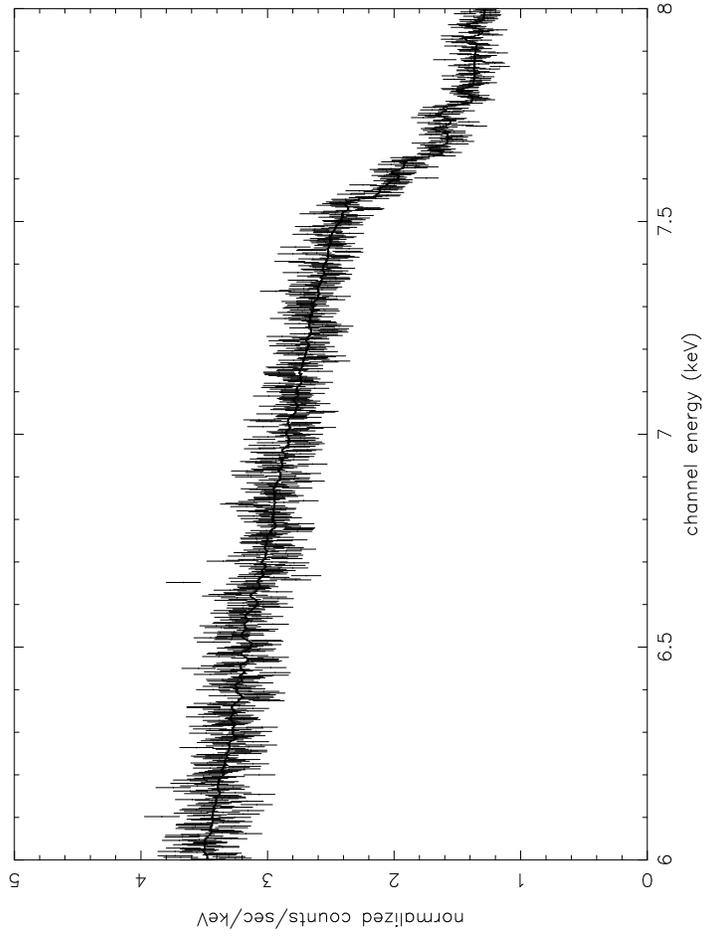}
\caption{Panel C:  Same as panel B, except using data from previous
versions of xstar (v2.1d)}
\end{figure}

\end{document}